\documentclass[prl,twocolumn,longbibliography]{revtex4-2}

\usepackage[colorlinks,urlcolor=blue,citecolor=blue,linkcolor=blue]{hyperref}

\usepackage{graphicx}
\usepackage{amsmath}
\usepackage{amssymb}
\usepackage{bm}
\usepackage{bbm}
\usepackage{mathtools}
\usepackage{graphics}
\usepackage{comment}
\usepackage{tikz}
\usepackage{mathtools}
\usepackage{physics}

\newcommand{\nn}{\nonumber}
\newcommand{\diff}{\mathop{}\!\mathrm{d}}

\renewcommand{\k}{\mathbf{k}}

\newcommand{\q}{\mathbf{q}}
\renewcommand{\r}{\mathbf{r}}

\newcommand{\Jhat}{\hat{J}}
\newcommand{\Hhat}{\hat{H}}
\newcommand{\tauhat}{\hat{\tau}}
\newcommand{\gammahat}{\hat{\gamma}}
\newcommand{\gammabold}{\hat{\bm{\gamma}}}
\newcommand{\epsilonbold}{\bm{\epsilon}}

\newcommand{\psiunder}{\underline{\psi}}

\DeclareMathOperator{\diag}{diag}
\DeclareMathOperator{\LK}{LK}
\DeclareMathOperator{\UT}{\hat{U}_T}
\DeclareMathOperator{\UTdagger}{\hat{U}_T^\dagger}
\DeclareMathOperator{\BdG}{BdG}
\DeclareMathOperator{\one}{\hat{\mathbbm{1}}}
\DeclareMathOperator{\transpose}{\mathrm{T}}

\newcommand{\clb}{\color{blue}}

\newcommand{\ignore}[1]{}

\begin{document}

\title{Collective modes in an unconventional superconductor with $j=3/2$ fermions}

\author{Guangyao Li}
\affiliation{Department of Physics and MacDiarmid Institute for Advanced Materials and Nanotechnology,
University of Otago, P.O. Box 56, Dunedin 9054, New Zealand}

\author{P. M. R. Brydon}
\affiliation{Department of Physics and MacDiarmid Institute for Advanced Materials and Nanotechnology,
University of Otago, P.O. Box 56, Dunedin 9054, New Zealand}

\begin{abstract}
    The $j = 3/2$ fermions in cubic crystals or cold atomic gases can form Cooper pairs in both singlet ($J = 0$) and unconventional quintet ($J = 2$) $s$-wave states. Our study utilizes analytical field theory to examine fluctuations in these states within the framework of the Luttinger-Kohn model. We investigate how collective modes evolve with varying spin-orbit coupling (SOC) strength. In the singlet state, quintet Bardasis-Schrieffer modes soften at a finite wavevector, hinting at Fulde-Ferrell-Larkin-Ovchinnikov physics. In the quintet state, we identify additional gapless and gapped modes originating from the partially broken symmetry due to SOC. Our results can be readily detected using current experimental techniques.
\end{abstract}

\maketitle

\emph{Introduction.}---While the existence of collective excitations in superconductors have been anticipated for many years~\cite{Parks}, only relatively recently have experimental advancements facilitated their detection~\cite{Col_mode_expment_1,Bohm_PRX_2014,Col_mode_expment_2,Col_mode_expment_3,REV_Higgs_mode}, prompting a surge of renewed interest in the field.  
The collective excitation spectrum in unconventional superconductors and superfluids is predicted to be particularly rich, reflecting the breaking of additional symmetries and their more intricate gap structure~\cite{SunPRR_2020,Qiao_PRB_2023,Poniatowski2022}. For example, the spin and orbital degrees of freedom in the Balian–Werthamer pairing state of superfluid $^3$He give rise to a complicated spectrum of collective modes~\cite{Sauls2022,Qiao_PRB_2023}.

A hallmark of unconventional superconductors is that their gap functions are nodal, leading to continuum excitations down to zero energy. The consequent damping of the collective modes by the continuum is unfavourable for their experimental detection. Recently there has been growing interest in unconventional pairing states with $s$-wave gap functions, which have been predicted to occur in systems where the low-energy fermionic states are characterized by quantum numbers beyond the usual spin-$\frac{1}{2}$, e.g. sublattice or spin~\cite{FuBerg2010,Ong2016,VafekChub,Kawakami2018,Nica2021}. In some materials such as YPtBi~\cite{YPtBi_experiment} or the pyrochlore iridates~\cite{Kondo2015}, the coupling between these different degrees of freedom give the band electron states an effective $j=3/2$. The pairing states of such systems have been extensively studied~\cite{Brydon_PRL_2016,Igor_PRL2018,Brydon_PRB_2018,Philip_PRB2019,Dutta_PRR_2021,Philip_PRB2023}.
The $j=3/2$ spin symmetry permits $s$-wave pairing states with $J=2$ (quintet) total angular momentum, making it a compelling system for investigating collective excitations in an unconventional superconductor.

\begin{figure}
    \centering
    \includegraphics[width=1\linewidth]{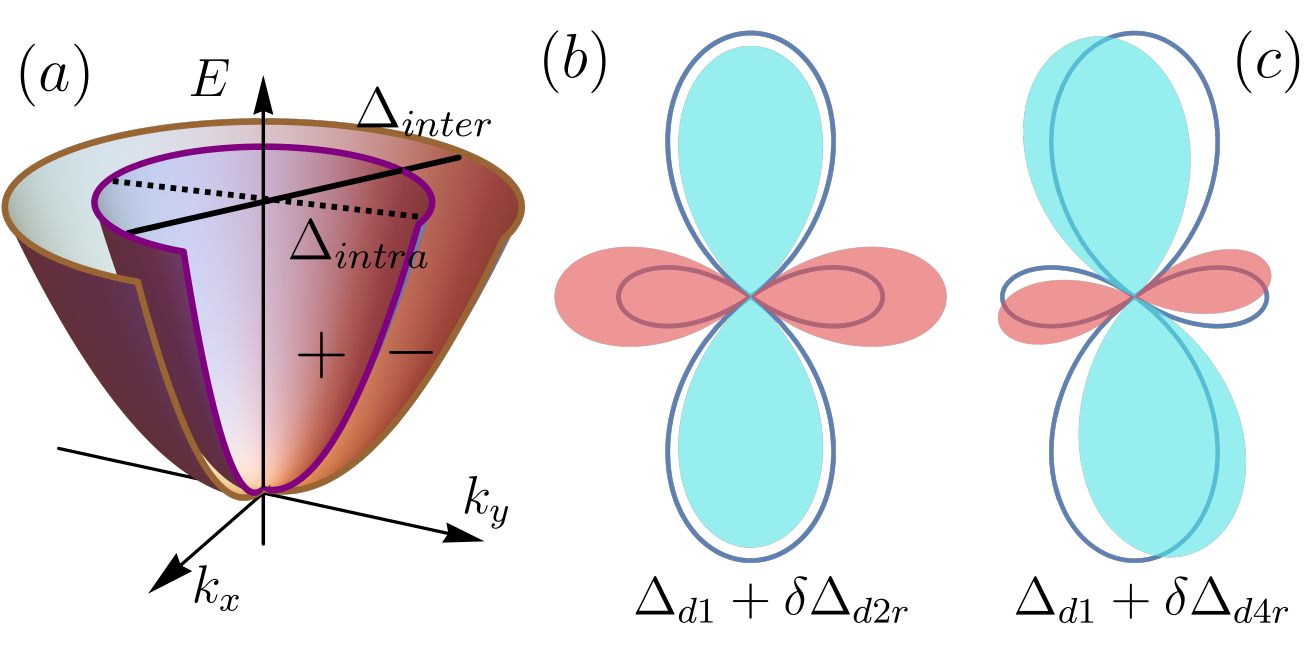}
    \caption{(a) Illustration of the split $\pm$ bands of $\hat{H}_{\text{LK}}$ at $k_z=0$. Quintet pairing involves both intraband and interband pairing. When the latter dominates, the system transitions to a finite momentum ground state, reflecting the FFLO physics, with SOC acting as an effective magnetic field. (b) Spherical harmonic representation of the fluctuation $\delta\Delta_{d2r}$ on top of the quintet pairing saddle point $\Delta_{d1}$. The presence of SOC changes the shape and size of the function and thus opens a gap.
(c) Fluctuation $\delta\Delta_{d4r}$ on top of the $\Delta_{d1}$ saddle point. The SOC serves as a rotation, and thus the excitation remains gapless.}
    \label{fig:harmonics}
\end{figure}

In this letter we explore the collective excitation modes in a $j=3/2$ system with quintet pairing and all symmetry-allowed spin-orbit coupling (SOC) terms. Utilizing a path integral approach, we derive an effective action for the gap fluctuations around saddle points corresponding to real and fully-gapped singlet ($J=0$) and quintet states. Our calculation scheme benefits from analytic expressions for the relevant Gor'kov Green's functions (GGF), which allows fast evaluation of the sums over momentum space. We first study the singlet saddle point, where we find that a pronounced softening of the quintet Bardasis-Schrieffer modes at nonzero wavevector is possible for sufficiently large SOC and interaction strengths. This indicates an instability towards a pairing state exhibiting a finite center-of-mass momentum, recalling Fulde-Ferrel-Larkin-Ovchinikov (FFLO) physics~\cite{FFLO_1,FFLO_2}. Instead of a magnetic field, the intrinsic interband and intraband pairing in the quintet state drives this phenomenon, see Fig.~\ref{fig:harmonics}(a).
The excitations about the quintet saddle point show a rich spectrum, with the appearance of massless amplitude modes corresponding to rotation of the nematic director of the real quintet state; these excitations, along with other distortions of the quintet state which are present as gapped modes, are shown in the cartoon Fig.~\ref{fig:harmonics}(b) and (c).
Our findings lay the groundwork for the study of the collective excitations in $j=3/2$ superconductors, which can be realized in cubic solid-state systems or as engineered Cooper pairs in cold-atomic gases.

\emph{Model.}---We consider a system of $j=3/2$ fermions with local pairing interactions, which is described by the Lagrangian:
\begin{align}
    L&=\int \diff\mathbf{r} \left[ \psiunder^\dagger \left(\partial_\tau +\hat{H}_{\LK}-\mu   \right) \psiunder +\mathcal{L}_s+ \mathcal{L}_d\right],
\end{align}
where $\psiunder=(c_{3/2}, c_{1/2}, c_{-1/2}, c_{-3/2})^{\transpose}$ is a spinor of Grassmann fields and $\tau$ is imaginary time.
The noninteracting part of the Lagrangian is the Luttinger-Kohn Hamiltonian, which is a minimal model for 
$j = 3/2$ fermions in a cubic material including SOC \cite{Luttinger_PR1955}. In momentum space this has the matrix form:
\begin{align}\label{eq:H_LK}
\Hhat_{\LK}=(\alpha \k^2-\mu)\one_4+\beta_J \sum_i k_i^2 \Jhat_i^2 +\delta_J \sum_{i\neq \mathrm{j}} k_i k_\mathrm{j} \Jhat_i \Jhat_\mathrm{j},
\end{align}
where $\one_4$ represents the $4 \times 4$ identity matrix and the  $\Jhat_{i,\mathrm{j}=x,y,z}$ are the $j=3/2$ angular momentum matrices.
The parameters $\alpha$, $\beta_J$, and $\delta_J$ are material-dependent quantities, and $\mu$ is the chemical potential.   If $\delta_J=\beta_J$ then $\Hhat_{\LK}$ is spherically symmetric; otherwise $\Hhat_{\LK}$ has cubic symmetry. In the following it is convenient to use the five mutually anticommuting Euclidean Dirac matrices $\{\gammahat_a\}_{a=1\ldots 5}$\cite{Igor_PRL2018,Philip_PRB2019,Philip_PRB2023}, which we define as follows: $\gammabold=(\frac{1}{3}(2\Jhat_z^2-\Jhat_x^2-\Jhat_y^2),\frac{1}{\sqrt{3}}(\Jhat_x^2-J_y^2), \frac{1}{\sqrt{3}}\{\Jhat_x,\Jhat_y\},  \frac{1}{\sqrt{3}}\{\Jhat_x,\Jhat_z\} , \frac{1}{\sqrt{3}}\{\Jhat_y,\Jhat_z\})$, where the inner curly brackets represent the anticommutation operation. Then 
$\Hhat_{\LK}$ can be written as $\Hhat_{\LK}=\epsilon_0\one_4+\sum_a \epsilon_a\gammahat_a$, where $\epsilon_0=(\alpha +\tfrac{5}{4}(\beta_J+\delta_J))\k^2-\mu$ and the vector of coefficients $\{\epsilon_a\}_{a=1\ldots 5}$ is $\epsilonbold=((\beta_J+\delta_J)(3k_z^2-\k^2)/2,\sqrt{3}(\beta_J+\delta_J)(k_x^2-k_y^2)/2, \sqrt{3}\beta_J k_x k_y, \sqrt{3}\beta_J k_x k_z, \sqrt{3}\beta_J k_y k_z )$. The doubly-degenerate eigenvalues of $\Hhat_{\LK}$ are given by $\epsilon_\pm=\epsilon_0\pm \abs{\epsilonbold}$, where $\abs{\epsilonbold}\equiv\sqrt{\sum_a \epsilon_a^2}$, see Supplemental Material (SM).

The effective spin-$3/2$ of the electrons allows $s$-wave pairing in both singlet ($J=0$) and quintet ($J=2$) channels~\cite{Igor_PRL2018,Brydon_PRL_2016}. To explore these pairing states we include the local pairing interaction terms
\begin{align}
\mathcal{L}_s&=-g_s  \left(\psiunder^\dagger  \UT \psiunder^* \right)\left(\psiunder^{\transpose}\UTdagger\psiunder   \right),\\
\mathcal{L}_d &=-g_d\sum_a  \left(\psiunder^\dagger \gammahat_a \UT \psiunder^* \right)\left(\psiunder^{\transpose}\UTdagger\gammahat_a \psiunder   \right),
\end{align}
where $g_s$ and $g_d$ represent the pairing strength in the
singlet and quintet channels respectively, and $\UT=\exp(-i\pi \Jhat_y)$ is the unitary part of the time reversal operator. 
We follow the standard procedure of decoupling the interactions $\mathcal{L}_s$ and $\mathcal{L}_d$ by introducing the bosonic fields $\Delta_s$ and $\{\Delta_{da}\}_{a=1\ldots 5}$ and performing the Hubbard-Stratanovich transformation~\cite{textbook_simons_2010}.
Upon integrating out the Grassmann fields (see SM), the partition function becomes:
\begin{align}
&\mathcal{Z}=\int\mathcal{D}(\Delta_s^*,\Delta_s)\int \prod_{a=1}^5 \mathcal{D}(\Delta_{da}^*,\Delta_{da})\,e^{-S},\\
&S=\int_0^\beta \diff \tau \sum_\k \left(\abs{\Delta_s}^2/g_s +\sum_{a=1}^5 \abs{\Delta_{da}}^2/g_d -\ln \det \hat{G}^{-1}  \right),
\end{align}
where $\beta=1/k_B T$ with $k_B$ the Boltzmann constant and $\hat{G}^{-1}$ represents the inverse of the GGF, which, following a Fourier transformation, is expressed as:
 \begin{align}\label{eq:Gorkov_G}
     \hat{G}^{-1}=\begin{pmatrix}
 i \omega_n \one_4-\Hhat_{\LK} & -\hat{\Delta}\\
 -\hat{\Delta}^\dagger & i \omega_n \one_4+\Hhat_{\LK}^{\transpose}
 \end{pmatrix}=i \omega_n \one_8-\Hhat_{\BdG}.
 \end{align}
 Here $\hat{\Delta} = (\Delta_s + \sum_a\Delta_{da}\hat{\gamma}_a)\UT$ and 
$\Hhat_{\BdG}$ is the Bogoliubov–de Gennes (BdG) Hamiltonian. In the following we will consider saddle points of this action where only the singlet or one of the quintet bosonic fields has a nonzero value.

\emph{Singlet saddle point.}---We first examine the collective excitations in the spin-singlet pairing state. The saddle-point value of the singlet gap is the solution of the gap equation $g_s^{-1} = \sum_{\k}(1/E_{+,\k}+1/E_{-,\k}) $, which we denote as $\Delta_{s0}$ and assume to be real, and $E_{\pm,\k}=\sqrt{\epsilon^2_{\pm}+\Delta^2_{s0}}$ are the eigenenergies of $\Hhat_{\BdG}$.
{Note that the summation over momentum in the gap equation requires a cutoff to converge, making the resulting expression an effective renormalization condition for the interaction strength $g_s$.} 
To study fluctuations about the saddle point we make the Ansatz 
$\Delta_s = \Delta_{s0}+\delta\Delta_s$ and $\Delta_{da}=\delta\Delta_{da}$, where the fluctuation terms have both real and imaginary components, e.g. $\delta\Delta_{s}=\delta\Delta_{sr}+i\delta\Delta_{si}$, representing amplitude and phase fluctuations, respectively. Expanding the action in the fluctuations to second order we obtain the Gaussian form:
\begin{align}
S=&\sum_{\omega_n,\q}\sum_{\nu=r,i}\left[g_s^{-1}+\chi_{s,\nu}(\omega_n,\q)  \right] \abs{\delta\Delta_{s\nu}}^2\notag\\
& + \sum_{\omega_n,\q}\sum_{\nu=r,i}\sum_a\left[g_d^{-1}+\chi_{da,\nu}(\omega_n,\q)  \right] \abs{\delta\Delta_{da\nu}}^2,\label{eq:action_expansion}
\end{align}
where the susceptibilities (or response functions) are:
\begin{equation}
    \chi_{\lambda,\nu}=\frac{1}{2\beta N}\sum_{\omega_m,\k}\Tr\left[\hat{M}_{\lambda,\nu} \hat{G}(\mathrm{k}) \hat{M}_{\lambda,\nu}  \hat{G}(\mathrm{k}+\mathrm{q})   \right].\label{eq:susceptibility_definition}
\end{equation}
Here $\hat{M}_{\lambda=s(da),\nu=r(i)}$ is the coefficient of the real (imaginary) part of the bosonic field $\Delta_{\lambda=s(da)}$ in the BdG Hamiltonian, and we adopt the abbreviation $\mathrm{k} = (\omega_n,\k)$ and $\mathrm{q}=(\omega_m,\q)$. Note that $N$ originates from lattice regularization and is sufficiently large to allow numerical approximations of momentum summation by integration.

The susceptibilities Eq.~\eqref{eq:susceptibility_definition} have a logarithmic divergence~\cite{SunPRR_2020}. In the case of the singlet channel this is precisely canceled by replacing $g_s^{-1}$ by the gap equation, leading to a universal function which is cut-off independent. 
This does not hold for the quintet fluctuations: although adding and subtracting $g_s^{-1}$ to the expression inside the brackets allows us to cancel the logarithmic divergence of the susceptibility, the solution then depends on the nonuniversal value of $g_d^{-1}-g_s^{-1}$~\cite{SunPRR_2020}.
In evaluating the regularized coefficients we are aided by the particularly simple analytical form of the GGF at the singlet saddle point:
\begin{align}
&\hat{G}_s=\sum_{\nu=\pm}\frac{-i\omega_n\one_8-  \epsilon_\nu \tauhat_z \otimes \one_4+i \Delta_{s0}\tauhat_y\otimes  \gammahat_3 \gammahat_5}{\omega_n^2+\epsilon_\nu^2 + \Delta_{s0}^2}\,\hat{\mathcal{P}}_\nu, \label{eq:projection_s_G}\\
& \hat{\mathcal{P}}_{\pm}=\frac{1}{2}\begin{bmatrix}
    \one_4\pm \sum_a \frac{{\epsilon}_a}{|\epsilonbold|}\hat{\gamma}_a & 0\\
    0 & \one_4 \pm \sum_a \frac{{\epsilon}_a}{|\epsilonbold|}\hat{\gamma}_a^{\transpose}
\end{bmatrix},\label{eq:projection_s_P}
\end{align}
where $\tauhat_{x,y,z}$ are Pauli matrices in the Nambu space and  $\hat{\mathcal{P}}_\pm$ is the projection operator into the band $\epsilon_\pm$.  We use Eq.~\eqref{eq:projection_s_G} to evaluate the coefficients in Eq.~\eqref{eq:action_expansion}, 
performing the Matsubara sum analytically and the integration over $\k$ numerically.  For the fluctuations in the singlet order parameter, we find the expected Higgs mode at the gap edge from the $\delta\Delta_{sr}$ fluctuations, while the $\delta\Delta_{si}$ fluctuation generates the massless Goldstone mode. Both modes are insensitive to the spherically symmetric SOC.

\begin{figure}
    \centering
    \includegraphics[width=1\linewidth]{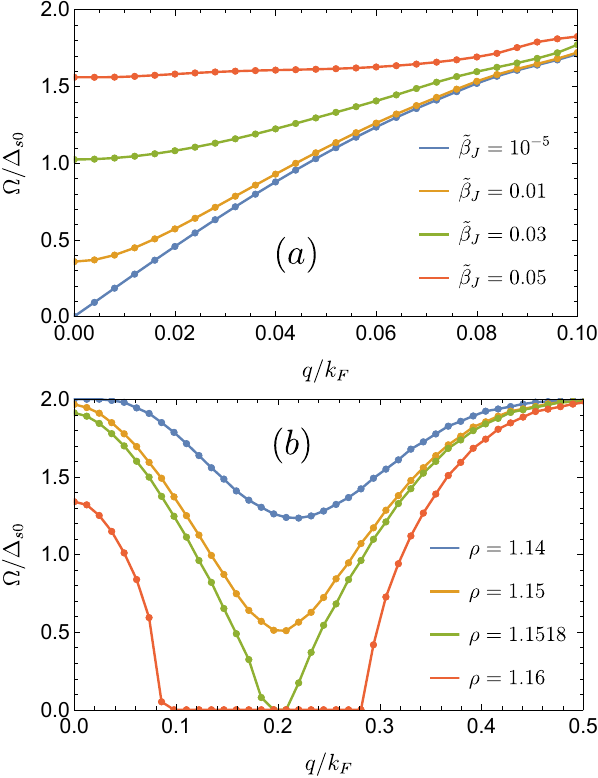}
    \caption{Dispersion of the Bardasis-Schrieffer mode of $\delta\Delta_{d1i}$ at $\Delta_{s0}/\mu=0.05$. (a) Evolution as a function of the dimentionless SOC parameter $\tilde{\beta}_J=\beta_J\, \mu/k_F^2\hbar^2$ with $\delta_J=\beta_J$ (spherical symmetric SOC) and $\rho=g_d/g_s=1$. (b) Evolution of instability with variations $\rho$, fixing $\beta_J/\Delta_{s0}=3.6$ and $\delta_J=-\beta_J$. Points represent numerical data, while solid curves serve as visual guides; $k_F$ is the Fermi momentum.}
    \label{fig:s_gap_dispersion}
\end{figure}

A more intriguing result emerges from the fluctuations into the quintet channel. To be concrete, Fig.~\ref{fig:s_gap_dispersion} illustrates the collective modes associated with $\delta\Delta_{d1i}$, corresponding to imaginary fluctuations in the $\ket{J,m_j}=\ket{2,0}$ state. In the absence of SOC and with equal pairing strength $\rho=g_d/g_s=1$, the degenerate singlet and quintet states give rise to an $\operatorname{SU}(4)$ symmetry~\cite{Wu_2003}; this is broken by selecting the singlet saddle point, thereby generating five additional Goldstone modes corresponding to phase fluctuations into the quintet channels.
The presence of a spherically symmetric SOC lifts this degeneracy: Given that the singlet pairing state opens the largest gap at the Fermi surface~\cite{Philip_PRB2023}, it is the ground state for $\rho=1$.
The quintet Goldstone modes found above are now gapped and can be regarded as Bardasis-Schrieffer modes~\cite{BardasisSchrieffer_1961}, i.e. fluctuations into a subdominant pairing state of different symmetry. As the strength of the SOC increases, the dispersion of these modes shifts upward toward the gap edge of $2\Delta_{s0}$, indicating the reduction of the effective interaction in the quintet channel. 
It is noteworthy that the top red curve in Fig.~\ref{fig:s_gap_dispersion}(a) develops a local minimum at a finite $q\equiv\abs{\q}$ value. This occurrence signals a mode softening process and may foreshadow a phase transition into a finite $q$ state.

To demonstrate this observation, 
Fig.~\ref{fig:s_gap_dispersion}(b) illustrates the phase transition process by adjusting $\rho$. As $\rho$ increases, the minimum value at finite $q$ may become lower than that at $q=0$. At the critical point, the excitation mode becomes gapless, signaling that the singlet saddle point is no longer the genuine ground state, and the system will transition to a pairing state at finite $q$. 
Physically, as the quintet channel involves both intra and interband pairing of $\hat{H}_{\LK}$ split by the SOC, the system transitions to a finite-$q$ state when the interband pairing dominates, as sketched in Fig.~\ref{fig:harmonics}(a). Consistent with this interpretation, we only observed the finite-$q$ state upon introducing a cubic anisotropy of the SOC which enhances the interband pairing in the $d1$ channel. 

This finite-$q$ instability is reminiscent of the FFLO state in spin-$\frac{1}{2}$ Pauli-limited singlet superconductors~\cite{FFLO_1,FFLO_2}. The interband pairing induced by the SOC plays the same pair-breaking role as the Zeeman field in a spin-$\frac{1}{2}$ singlet superconductor, in clear analogy to FFLO physics. A key difference is that in general there is also intraband pairing for any cubic anisotropy of the SOC, which in the analogy to the spin-$\frac{1}{2}$ superconductor corresponds to a same-spin pairing component. This is insensitive to the Zeeman field, but in general it cannot coexist with singlet pairing. The presence of intraband pairing means that the $q=0$ quintet state remains a weak-coupling instability for any cubic anisotropy~\cite{Brydon_PRB_2018}, although our results here show that it may be unstable towards a finite-$q$ state.

\begin{figure}
    \centering
    \includegraphics[width=1\linewidth]{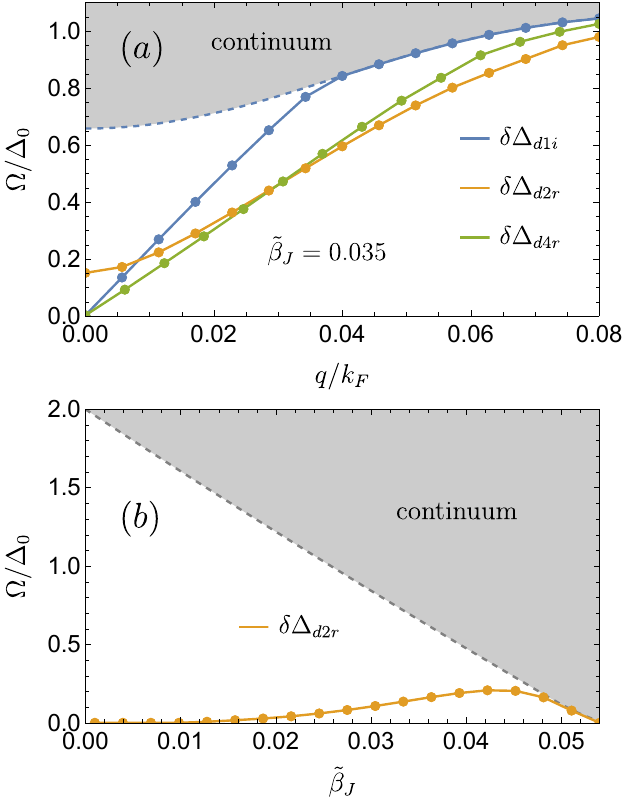}
    \caption{Excitation spectrum (solid curves) at the $\Delta_{d1}$ saddle point with $\Delta_0/\mu=0.05$. (a) Comparative analysis of various fluctuation channels at $\tilde{\beta}_J=\beta_J \mu/k_F^2\hbar^2=0.035$. The blue dashed curve denotes the effective gap. (b) Fluctuations on the $\delta\Delta_{d2r}$ channel with respect to $\tilde{\beta}_J$ at $q=0$. The grey dashed curve illustrates the reduced gap. In both panels, the shaded grey area represents the continuum.}
    \label{fig:d1_gap}
\end{figure}

\emph{Quintet saddle point.}---We now apply Eq.~\eqref{eq:action_expansion} to probe the physics of the quintet saddle point. 
The five quintet states are degenerate in the absence of SOC, and the saddle point solution corresponds to any real combination of the different channels~\cite{HoYip_PRL_1999}. Real solutions remain stable and fully-gapped for sufficiently small but finite SOC~\cite{Philip_PRB2019}, although they display gap minima which deepen into line nodes at some critical SOC strength. Within the manifold of real solutions, switching on the SOC splits off the uniaxial nematic $\ket{2,0}$ ($d1$) state from the four degenerate biaxial nematic states ($d2$-$d5$)~\cite{Herbut_PRB_2019}; we expect that the $\ket{2,0}$ state is most stable since the gap minima form nonintersecting circles, in contrast to the intersecting lines of gap minima in the other states. 

The saddle point equation is $\delta S|_{g_s=0}/\delta \Delta_{d1}=0$, and its solution is denoted as $\Delta_0$. We proceed as before by expanding to quadratic order in the fluctuations. Here the coefficients of the quintet fluctuations are universal, since the saddle point equation allows us to replace $g_d^{-1}$ by a term which cancels the logarithmic divergence of the susceptibility. We again use an analytic expression for the GGF to exactly perform the Matsubara sum, although this is much more complicated than at the singlet saddle point and we present it in the SM. In the following we ignore the singlet channel which in general contributes a Bardasis-Schrieffer mode.

At vanishing SOC strength we find five massless modes: the usual Goldstone mode due to the $\operatorname{U}(1)$ symmetry breaking, 
and four additional degenerate amplitude modes arising from the breaking of $\operatorname{SO}(5)$ symmetry within the five-dimensional order parameter space~\cite{Wu_2003}. As shown in Fig.~\ref{fig:d1_gap}(a), three of these Goldstone modes survive upon switching on the spherically-symmetric SOC: the phase fluctuation mode associated with $\delta\Delta_{d1i}$, and the two amplitude modes associated with $\delta\Delta_{d4r}$ and $\delta\Delta_{d5r}$. The amplitude modes $\delta\Delta_{d2r}$ and $\delta\Delta_{d3r}$ acquire a small mass gap, and move towards the edge of the continuum with increasing SOC as shown in Fig.~\ref{fig:d1_gap}(b).

The continued presence of Goldstone modes at finite SOC strength follows from the breaking of the $\operatorname{SO}(3)$ symmetry by the nematic director of the $d1$ state. To see the effect of this, let us consider an infinitesimal rotation about the $y$-axis, $\hat{U}_y = \exp(-i \phi_y \hat{J}_y)$, where we have set $\hbar=1$. As a result of this transformation, $\k\mapsto\k^\prime$ and $\hat{H}_{\BdG}$ undergoes the following changes:
$\Hhat_{\BdG}(\k^\prime)=\diag(\hat{U}_y^\dagger, \hat{U}_y^*)\Hhat_{\BdG}(\k)\diag(\hat{U}_y,\hat{U}_y^{\transpose})$. To the lowest nontrivial order in $\phi_y$, the pairing potential transforms as:
\begin{align}
    \hat{U}_y^\dagger \hat{\Delta}_{d1}(\k)\hat{U}_y^{\transpose}=\hat{\Delta}_{d1}(\k^\prime)-\Delta_0\phi_y \gammahat_4 \UT + \mathcal{O}(\phi_y^2).
\end{align}
This implies that the infinitesimal rotation of $\hat{H}_{\BdG}$ about the $y$-axis corresponds to amplitude fluctuations into the $\delta\Delta_{d4}$ channel ($\frac{1}{2i}(\ket{2,1}-\ket{2,-1})$ state). Owing to the summation over momentum, the total energy of the system remains unchanged by this rotation. Consequently, the Goldstone theorem dictates the existence of a massless mode corresponding to the $\delta\Delta_{d4r}$ channel at finite strength of SOC. A similar argument can be made for the rotation against the $x$-axis and the fluctuation of the $\delta\Delta_{d5r}$ channel ($\frac{1}{2}(\ket{2,1}+\ket{2,-1})$ state). The $\delta\Delta_{d4r}$ and $\delta\Delta_{d5r}$ fluctuations remain degenerate upon including a finite cubic anisotropy but the absence of $\operatorname{SO}(3)$ rotational symmetry in the normal state implies a nonzero mass for these modes.

The distinct behavior of fluctuations in the non-pairing channels can be illustrated using a heuristic cartoon picture depicted in Fig.~\ref{fig:harmonics}, which visualizes the quintet pairing states as $L=2$ spherical harmonic functions in real space. The $\delta\Delta_{d2r}$ fluctuations corresponds to a change in the shape of the harmonic function, thus requiring a finite amount of energy. In contrast, the $\delta\Delta_{d4r}$ fluctuations corresponds to an overall rotation, and so this fluctuation requires no energy.
The symmetry-breaking process induced by the $\delta\Delta_{d2r}$ channel is clearly evident in the evolution of the $q=0$ frequency against SOC strength, as depicted in Fig.~\ref{fig:d1_gap}(b). The mass gap increases with $\beta_J$, but this trend appears to reverse near to the closing of the excitation gap.

\emph{Summary.}---In this letter we have presented a study of the collective modes in fully-gapped pairing states in a system of $j=3/2$ fermions. Specializing to the real singlet and quintet saddle points, we find a rich diversity of collective modes. At 
the singlet saddle point, our investigation reveals that interband pairing may induce a mode softening process which can disrupt the assumed singlet ground state. This phenomenon resembles the FFLO physics but without the requirement of a high external magnetic field. Consequently, the quintet system offers a unique platform for investigating FFLO physics, and reciprocally, the FFLO state may unexpectedly give insight into the quintet system with SOC.
Exploring the quintet saddle point, we observe that in the presence of SOC, the breaking of rotational symmetry by the uniaxial nematic phases gives rise to two additional gapless amplitude modes alongside the usual phase mode.
Observation of these collective modes is feasible in either cubic superconductors or in cold atomic gases of quintet Cooper pairs using current experimental techniques. 
Our results underline the important role that SOC plays in unconventional multiband superconductors.

\begin{acknowledgments}
We thank Daniel Agterberg for useful discussions. This work was supported by
the Marsden Fund Council from Government funding, managed by Royal Society Te Ap\={a}rangi, Contract No. UOO1836.
\end{acknowledgments}

\onecolumngrid
\appendix

\section{Spin $j=3/2$ Angular Momentum Operator and $\hat{\gamma}_a$ Matrices}

The matrix representation of the $j=3/2$ angular momentum operators is given by:
\begin{align}		
\Jhat_x=\frac{1}{2}\begin{bmatrix}
	0 & \sqrt{3} & 0 & 0 \\
	\sqrt{3} & 0 & 2 & 0 \\
	0 & 2 & 0 & \sqrt{3} \\
	0 & 0 & \sqrt{3} & 0 \\
\end{bmatrix}, \qquad \Jhat_y=\frac{i}{2}\begin{bmatrix}
0 & -\sqrt{3} & 0 & 0 \\
\sqrt{3} & 0 & -2 & 0 \\
0 & 2 & 0 & -\sqrt{3} \\
0 & 0 & \sqrt{3} & 0 \\
\end{bmatrix}, \qquad \Jhat_z=\frac{1}{2}\begin{bmatrix}
3 & 0 & 0 & 0 \\
0 & 1 & 0 & 0 \\
0 & 0 & -1 & 0 \\
0 & 0 & 0 & -3 \\
\end{bmatrix}.
\end{align}

The $\gammahat_a$ matrices are:
\begin{align}
	&\gammahat_1=\frac{1}{3}\left(2\Jhat_z^2-\Jhat_x^2-\Jhat_y^2\right)=\begin{bmatrix}
		1 & 0 & 0 & 0 \\
		0 & -1 & 0 & 0 \\
		0 & 0 & -1 & 0 \\
		0 & 0 & 0 & 1 \\
	\end{bmatrix}, \qquad \gammahat_2=\frac{1}{\sqrt{3}}\left(\Jhat_x^2-J_y^2\right)=\begin{bmatrix}
	0 & 0 & 1 & 0 \\
	0 & 0 & 0 & 1 \\
	1 & 0 & 0 & 0 \\
	0 & 1 & 0 & 0 \\
	\end{bmatrix},\\[1em]
	&\gammahat_3=\frac{1}{\sqrt{3}}\{\Jhat_x,\Jhat_y\}=\begin{bmatrix}
		0 & 0 & -i & 0 \\
		0 & 0 & 0 & -i \\
		i & 0 & 0 & 0 \\
		0 & i & 0 & 0 \\
	\end{bmatrix}, \quad \gammahat_4=\frac{1}{\sqrt{3}}\{\Jhat_x,\Jhat_z\}=\begin{bmatrix}
	0 & 1 & 0 & 0 \\
	1 & 0 & 0 & 0 \\
	0 & 0 & 0 & -1 \\
	0 & 0 & -1 & 0 \\
	\end{bmatrix},\quad
	\gammahat_5=\frac{1}{\sqrt{3}}\{\Jhat_y,\Jhat_z\}=\begin{bmatrix}
		0 & -i & 0 & 0 \\
		i & 0 & 0 & 0 \\
		0 & 0 & 0 & i \\
		0 & 0 & -i & 0 \\
	\end{bmatrix}.
\end{align}
The unitary component of the time reversal operator is given by:
\begin{equation}
	\UT=\exp\left(-i\pi \Jhat_y\right)=\begin{bmatrix}
		0 & 0 & 0 & -1 \\
		0 & 0 & 1 & 0 \\
		0 & -1 & 0 & 0 \\
		1 & 0 & 0 & 0 \\
	\end{bmatrix}.
\end{equation}

\section{path integral representation for the partition function}

The Luttinger-Kohn (LK) Hamiltonian $\Hhat_{\LK}$ defined in the main text reads: 
\begin{align}\label{eq:H_LK}
	\Hhat_{\LK}=(\alpha \k^2-\mu)\one_4+\beta_J \sum_i k_i^2 \Jhat_i^2 +\delta_J \sum_{i\neq j} k_i k_j \Jhat_i \Jhat_j.
\end{align}
Utilizing the $\gammahat_a$ matrices and the $\epsilonbold$ vector, the doubly-degenerate eigenvalues of $\Hhat_{\LK}$ are: $\epsilon_\pm=\epsilon_0+\sqrt{\sum_a \epsilon_a^2}$.

We consider an ensemble of non-interacting electrons described by $\Hhat_{\LK}$ as the the non-interacting Hamiltonian $H_0$. In terms of the electron creation and annihilation operators base vector $\hat{\psi}=(\hat{c}_{3/2}, \hat{c}_{1/2}, \hat{c}_{-1/2}, \hat{c}_{-3/2})^{\transpose}$, $H_0$ is given by:
\begin{align}
	H_0=\int\diff\r\,\left[ \hat{\psi}^\dagger \Hhat_{\LK} \hat{\psi}\right].
\end{align}
Next, we introduce the interaction Hamiltonian $H_I$, defined as:
\begin{align}
	H_I=\int\diff\r\,\left[-g_s  \left(\hat{\psi}^\dagger  \UT \hat{\psi}^* \right)\left(\hat{\psi}^{\transpose}\UTdagger\hat{\psi}\right)  -g_d\sum_a  \left(\hat{\psi}^\dagger \gammahat_a \UT \hat{\psi}^* \right)\left(\hat{\psi}^{\transpose}\UTdagger\gammahat_a \hat{\psi}\right)  \right].
\end{align}
The partition function of the full system is: $\mathcal{Z}=\tr e^{-\beta(H_0+H_I)}$. By inserting the resolution of identity of the coherent states as outlined in Ref.~\cite{textbook_simons_2010}, we arrive at the path integral description of the partition function written against the Grassmann field basis $\psiunder=(c_{3/2}, c_{1/2}, c_{-1/2}, c_{-3/2})^{\transpose}$ as:
\begin{align}
	\mathcal{Z}&=\int D(\psiunder^\dagger, \psiunder) \exp\left\{  -\int_0^\beta\diff\tau\int\diff\r \left[\psiunder^\dagger\left(\partial_\tau+\Hhat_{\LK}-\mu\right)\psiunder + \mathcal{L}_s + \mathcal{L}_d \right]\right\}, \nn\\[1em]
	\mathcal{L}_s&= -g_s  \left(\psiunder^\dagger  \UT \psiunder^* \right)\left(\psiunder^{\transpose}\UTdagger\psiunder\right), \nn \\
	\mathcal{L}_d&=-g_d\sum_a  \left(\psiunder^\dagger \gammahat_a \UT \psiunder^* \right)\left(\psiunder^{\transpose}\UTdagger\gammahat_a \psiunder\right).  
\end{align}
The quartic interaction appearing in the action does not allow analytic integration, but it can be transformed away by using the Hubbard-Stratonovich transformation which introduces another bosonic field as a dynamic variable. Take $\mathcal{L}_s$ as an example.
To perform the Hubbard-Stratonovich transformation to cancel out the quartic interaction, we can multiply $\mathcal{Z}$ by a constant, which is expressed as a functional integration of a bosonic field and its conjugate:
\begin{align}
	\int D(\Delta_s^*, \Delta_s) e^{-\int\diff \tau \int\r\, \frac{\abs{\Delta_s}^2}{g_s}}.
\end{align}
Under this operation, all thermodynamic variables will remain unchanged  as they are given by the derivative of $\ln \mathcal{Z}$ such that the additional constant will cancel out itself between the numerator and the denominator. We can further manipulate the constant by making a shift to the bosonic field as:
\begin{align}
    \Delta_s \to \Delta_s -g_s \psiunder^{\dagger}\UT\psiunder^*,\quad \text{and}\quad
      	\Delta_s^* \to \Delta_s^* -g_s \psiunder^{\transpose}\UTdagger\psiunder.
\end{align}
The product of the shifted fields in $\abs{\Delta_s}^2/g_s$ will cancel out the quartic term in $\mathcal{L}_s$. Similar argument applies to the $\mathcal{L}_d$ term. The remaining terms is quadratic in the Grassmann fields and can be integrated out by the Gaussian integral \cite{textbook_simons_2010}, and the resulting partition function expressed by functional integration of bosonic field is:
\begin{align}
	\mathcal{Z}=\int\mathcal{D}(\Delta_s^*,\Delta_s)\int \prod_{a=1}^5 \mathcal{D}(\Delta_{da}^*,\Delta_{da})\,\exp\left\{- \int_0^\beta \diff \tau \sum_\k \left( \frac{\abs{\Delta_s}^2}{g_s} +\sum_{a=1}^5 \frac{\abs{\Delta_{da}}^2}{g_d} -\ln \det \hat{G}^{-1}  \right)    \right\}.
\end{align}
This expression serves as the central model in our discussions in the main text.

\section{spin-singlet pairing state}

The spin-singlet pairing state corresponds to the singlet saddle point solution. On the singlet saddle point, we can write down the meanfield Bogoliubov–de Gennes (BdG) Hamiltonian in the Nambu basis as:
\begin{align}
	\Hhat_{\BdG,\k}=\begin{bmatrix}
		\Hhat_{\LK} & \hat{\Delta}_s\\
		\hat{\Delta}_s^\dagger & -\Hhat_{\LK}^{\transpose}
	\end{bmatrix},
\end{align}
where the pairing potential is given by $\hat{\Delta}_s=-\Delta_{s0} \UT$. The positive eigenvalues of the BdG Hamiltonian are:
\begin{align}
	E_{\pm,\k} = \sqrt{\epsilon_{\pm}^2 + \Delta_{s0}^2}.
\end{align}

We find the singlet saddle point by solving the saddle point equation and then expand the action in terms of the real and imaginary parts of the singlet channel fluctuation, as described in the main text. The expanded action and susceptibility (or response functions) are given by:
\begin{align}
S&=\sum_{\omega_n,\q}\sum_{\nu=r,i}\left[g_s^{-1}+\chi_{s,\nu}(\omega_n,\q)  \right] \abs{\delta\Delta_{s\nu}}^2,\label{eq:action_expansion}\\
    \chi_{s,\nu}&=\frac{1}{2\beta N}\sum_{\omega_m,\k}\Tr\left[\hat{M}_{s,\nu} \hat{G}(\omega_n,\k) \hat{M}_{s,\nu}  \hat{G}(\omega_n+\omega_m,\k+\q)   \right].\label{eq:susceptibility_definition}
\end{align}
Here, $\hat{M}_{s,\nu}$ is the off-diagonal block matrix of $\Hhat_{\BdG}$ following complex decomposition. $N$ comes from lattice regularization and it is assumed to be sufficiently large so that in numerical calculations, the summation over momentum can be approximated by integration.

To simplify expressions, from now on we omit the overhead hat symbol representing operators or matrices, and define $\abs{\epsilonbold}=\sqrt{\sum_a \epsilon_a^2}$. 
The Green's function has the form
\begin{eqnarray}
G_s(\k,i\omega_n) & = & (i\omega_n \mathbbm{1}_8 - H_{\BdG,\k})^{-1} \notag \\
& = & \frac{1}{(\omega_n^2 + E_{+}^2)(\omega_n^2 + E_{-}^2)}\left\{-i\omega_n(\Delta_0^2 + \epsilon_0^2 +
\abs{\epsilonbold}^2 + \omega_n^2)\tau_0\mathbbm{1}_4 - \epsilon_0(\Delta_0^2 + \epsilon_0^2 -
\abs{\epsilonbold}^2 + \omega_n^2)\tau_z\mathbbm{1}_4\right. \notag
\\
&& +2i\omega_n\epsilon_0\epsilon_1\tau_0\gamma_1 -\epsilon_1(\Delta_0^2 - \epsilon_0^2 + \abs{\epsilonbold}^2 +
\omega_n^2)\tau_z\gamma_1 +2i\omega_n\epsilon_0\epsilon_2\tau_0\gamma_2 
-\epsilon_2(\Delta_0^2 - \epsilon_0^2 + \abs{\epsilonbold}^2 +
\omega_n^2)\tau_z\gamma_2 \notag \\
&&-\epsilon_3(\Delta_0^2 - \epsilon_0^2 + \abs{\epsilonbold}^2 +
\omega_n^2)\tau_0\gamma_3   + 2\Delta_0\epsilon_0\epsilon_5\tau_x\gamma_3 +
2i\omega_n\epsilon_0\epsilon_3\tau_z\gamma_3 +
2i\omega_n\epsilon_0\epsilon_4\tau_0\gamma_4\notag \\
&& -\epsilon_4(\Delta_0^2 - \epsilon_0^2 + \abs{\epsilonbold}^2 +
\omega_n^2)\tau_z\gamma_4  -\epsilon_5(\Delta_0^2 - \epsilon_0^2 + \abs{\epsilonbold}^2 +
\omega_n^2)\tau_0\gamma_5 -2\Delta_0\epsilon_0\epsilon_3\tau_x\gamma_5
+ 2i\omega_n\epsilon_0\epsilon_5\tau_z\gamma_5 \notag \\
&& +
2\Delta_0\epsilon_0\epsilon_4\tau_y i\gamma_1\gamma_2  -
2\Delta_0\epsilon_0\epsilon_2 \tau_y i\gamma_1\gamma_4 + 2\Delta_0\epsilon_0\epsilon_1\tau_y i\gamma_2\gamma_4 +
\Delta_0(\Delta_0^2 + \epsilon_0^2 + \abs{\epsilonbold}^2 +
\omega_n^2)\tau_y i\gamma_3\gamma_5\big{\}}\\
& = & \frac{1}{2(\omega_n^2 +
	E_+^2)}\left\{-i\omega_n\tau_0\mathbbm{1}_4 -
\epsilon_+\tau_z\mathbbm{1}_4 -
i\omega_n\epsilon_1\tau_0\gamma_1 -
\epsilon_1\epsilon_+\tau_z\gamma_1 -
i\omega_n\epsilon_2\tau_0\gamma_2 -
\epsilon_2\epsilon_+\tau_z\gamma_2\right. \notag \\
&& -i\omega_n\epsilon_3\tau_z\gamma_3 -
\epsilon_3\epsilon_+\tau_0\gamma_3 -i\omega_n\epsilon_4\tau_0\gamma_4 -
\epsilon_1\epsilon_+\tau_z\gamma_4-i\omega_n\epsilon_5\tau_z\gamma_5 -
\epsilon_5\epsilon_+\tau_0\gamma_5 +\Delta_0\tau_y
i\gamma_3\gamma_5\notag\\
&& \left. -\Delta_0\epsilon_1\tau_y i\gamma_2\gamma_4 +
\Delta_0\epsilon_2\tau_y i\gamma_1\gamma_4 +
\Delta_0\epsilon_3\tau_x \gamma_5 -
\Delta_0\epsilon_4\tau_y i\gamma_1\gamma_2
-\Delta_0\epsilon_5\tau_x \gamma_3\right\}\notag \\
&& + \frac{1}{2(\omega_n^2 +
	E_-^2)}\left\{-i\omega_n\tau_0\mathbbm{1}_4 -
\epsilon_-\tau_z\mathbbm{1}_4 +
i\omega_n\epsilon_1\tau_0\gamma_1 +
\epsilon_1\epsilon_-\tau_z\gamma_1 +
i\omega_n\epsilon_2\tau_0\gamma_2 +
\epsilon_2\epsilon_-\tau_z\gamma_2\right. \notag \\
&& +i\omega_n\epsilon_3\tau_z\gamma_3 +
\epsilon_3\epsilon_-\tau_0\gamma_3 +i\omega_n\epsilon_4\tau_0\gamma_4 +
\epsilon_4\epsilon_-\tau_z\gamma_4+i\omega_n\epsilon_5\tau_z\gamma_5 +
\epsilon_5\epsilon_-\tau_0\gamma_5 +\Delta_0\tau_y
i\gamma_3\gamma_5\notag\\
&& \left. -\Delta_0\epsilon_1\tau_y i\gamma_2\gamma_4 +
\Delta_0\epsilon_2\tau_y i\gamma_1\gamma_4 +
\Delta_0\epsilon_3\tau_x \gamma_5 -
\Delta_0\epsilon_4\tau_y i\gamma_1\gamma_2
-\Delta_0\epsilon_5\tau_x \gamma_3\right\} \notag \\
&= & \frac{1}{(\omega_n^2 +
	E_+^2)}(-i\omega_n\tau_0\mathbbm{1}_4 - \epsilon_+\tau_z\mathbbm{1}_4
+ \Delta_0\tau_yi\gamma_3\gamma_5)\frac{\left(\tau_0\mathbbm{1}_4 +
	\epsilon_1\tau_0\gamma_1 + \epsilon_2\tau_0\gamma_2 +
	\epsilon_3\tau_z\gamma_3 + \epsilon_4\tau_0\gamma_4 +
	\epsilon_5\tau_z\gamma_5\right)}{2} \notag \\
&& + \frac{1}{(\omega_n^2 +
	E_-^2)}(-i\omega_n\tau_0\mathbbm{1}_4 - \epsilon_-\tau_z\mathbbm{1}_4
+ \Delta_0\tau_yi\gamma_3\gamma_5)\frac{\left(\tau_0\mathbbm{1}_4 -
	\epsilon_1\tau_0\gamma_1 - \epsilon_2\tau_0\gamma_2 -
	\epsilon_3\tau_z\gamma_3 - \epsilon_4\tau_0\gamma_4 -
	\epsilon_5\tau_z\gamma_5\right)}{2} \notag \\
& = & \frac{1}{(\omega_n^2 +
	E_+^2)}(-i\omega_n\tau_0\mathbbm{1}_4 - \epsilon_+\tau_z\mathbbm{1}_4
+ \Delta_0\tau_yi\gamma_3\gamma_5)\begin{pmatrix}{\cal P}_{+} & 0 \\
	0 & {\cal P}^{(h)}_{+}\end{pmatrix} \notag \\
&& + \frac{1}{(\omega_n^2 +
	E_-^2)}(-i\omega_n\tau_0\mathbbm{1}_4 - \epsilon_-\tau_z\mathbbm{1}_4
+ \Delta_0\tau_yi\gamma_3\gamma_5)\begin{pmatrix}{\cal P}_{-} & 0 \\
	0 & {\cal P}^{(h)}_{-}\end{pmatrix}
\end{eqnarray}
where $\tau_{0xyz}$ are Pauli matrices that have been introduced to encode the Nambu degree of freedom. The projection operators for the electron and hole bands are given by :
\begin{eqnarray}
{\cal P}_\pm  & = & \frac{1}{2}\left[\mathbbm{1}_4 \pm \frac{\epsilonbold(\k)\cdot\gammabold}{\abs{\epsilonbold}}
\right],\\
{\cal P}^{(h)}_\pm & = & \frac{1}{2}\left[\mathbbm{1}_4 \pm \frac{\epsilonbold(-\k)\cdot\gammabold}{\abs{\epsilonbold}}
\right]^{\transpose}.
\end{eqnarray}
The projection operators allow us to perform analytic summation over the Matsubara frequency, which is done by Mathematica. After that, the divergence in the susceptibility can be canceled out by the pairing strength renormalized by the gap equation. The resulting expression is a well-behave function in both $\k$ and $\q$, and the plots in the main texts are calculated by numerical integration over $\k$, which converges very quickly.

\section{Spin quintet pairing}

In this section, we consider a spin-quintet pairing state, which without loss of
generality we choose to be 
\begin{equation}
\Delta = \Delta_0\, \gamma_1 \operatorname{U_T}
\end{equation}
The positive eigenvalues of the BdG Hamiltonian are 
\begin{equation}
E_{\pm} = \sqrt{\Delta_0^2 + \epsilon_0^2 +
	\abs{\epsilonbold}^2 \pm 2\sqrt{\Delta_0^2(\epsilon_2^2 + \epsilon_3^2
		+ \epsilon_4^2+
		\epsilon_5^2) + \epsilon_0^2\abs{\epsilonbold}^2}}\,.
\end{equation}
An important feature is that the eigenvalues $E_{\pm}$ do not directly map to
$\epsilon_\pm$ in the limit $\Delta_0\rightarrow0$. Rather, since
$E_{-}\leq E_{+}$, we have
\begin{equation}
E_{-} \rightarrow \text{min}\{|\epsilon_+|,|\epsilon_-|\}\,, \qquad E_{+} \rightarrow \text{max}\{|\epsilon_+|,|\epsilon_-|\}
\end{equation}
Thus, in contrast to the spin singlet pairing, here $E_{-}$ and
$E_{+}$ refer to ``low-energy'' and ``high-energy'' sectors, respectively. 

The Green's function has the form
\begin{eqnarray}
{\cal G}({\bf k},i\omega_n) & = & (i\omega_n \mathbbm{1}_8 - H_{\BdG,\k})^{-1} \notag \\
& = & \frac{1}{(\omega_n^2 + E_{+}^2)(\omega_n^2 + E_{-}^2)}\left\{-i\omega_n(\Delta_0^2 + \epsilon_0^2 +
\abs{\epsilonbold}^2 + \omega_n^2)\tau_0\mathbbm{1}_4 - \epsilon_0(\Delta_0^2 + \epsilon_0^2 -
\abs{\epsilonbold}^2 + \omega_n^2)\tau_z\mathbbm{1}_4\right. \notag \\
&& + 2i\omega_n\epsilon_0\epsilon_1\tau_0\gamma_1
-\epsilon_1(\Delta_0^2-\epsilon_0^2+\abs{\epsilonbold}^2+\omega_n^2)\tau_z\gamma_1
+2i\omega_n\epsilon_0\epsilon_2\tau_0\gamma_2 -
2i\omega_n\Delta_0\epsilon_4\tau_x\gamma_2 \notag \\
&& -\epsilon_2(-\Delta_0^2
-\epsilon_0^2+\abs{\epsilonbold}^2+\omega_n^2)\tau_z\gamma_2
-\epsilon_3(-\Delta_0^2
-\epsilon_0^2+\abs{\epsilonbold}^2+\omega_n^2)\tau_0\gamma_3
-2\Delta_0\epsilon_1\epsilon_5\tau_x\gamma_3 +
2i\omega_n\epsilon_0\epsilon_3\tau_z\gamma_3\notag \\
&& +2i\omega_n\epsilon_0\epsilon_4\tau_0\gamma_4 +
2i\omega_n\Delta_0\epsilon_2\tau_x\gamma_4 - \epsilon_4(-\Delta_0^2 -
\epsilon_0^2+\abs{\epsilonbold}^2+\omega_n^2)\tau_z\gamma_4 -
\epsilon_5(-\Delta_0^2 -
\epsilon_0^2+\abs{\epsilonbold}^2+\omega_n^2)\tau_0\gamma_5\nn\\
&& +2\Delta_0\epsilon_1\epsilon_3\tau_x\gamma_5 +
2i\omega_n\epsilon_0\epsilon_5\tau_z\gamma_5
-2\Delta_0\epsilon_1\epsilon_4\tau_yi\gamma_1\gamma_2 -
2i\omega_n\Delta_0\epsilon_5\tau_yi\gamma_1\gamma_3 +
2\Delta_0\epsilon_1\epsilon_2\tau_yi\gamma_1\gamma_4\notag \nn\\
&& \left.+ 2i\omega_n\Delta_0\epsilon_3\tau_yi\gamma_1\gamma_5 -
\Delta_0(\Delta_0^2 + \epsilon_0^2 +2\epsilon_1^2 - \abs{\epsilonbold}^2
+ \omega_n^2)\tau_yi\gamma_2\gamma_4 -2\Delta_0\epsilon_0\epsilon_1\tau_yi\gamma_3\gamma_5\right\}.
\end{eqnarray}
It is convenient to split the Green's function into a low-energy and a
high-energy part, i.e. we write
\begin{equation}
{\cal G}({\bf k},i\omega_n) = {\cal G}_{+}({\bf k},i\omega_n) + {\cal G}_{-}({\bf k},i\omega_n)
\end{equation}
where
\begin{eqnarray}
{\cal G}_{{\clb{\pm}}}({\bf k},i\omega_n) & = & \frac{1}{2(\omega_n^2 +
	E_{{\clb \pm}}^2)}\left\{ -i\omega_n\tau_0\mathbbm{1}_4 - \epsilon_0\left(1 \,{\clb \pm}\,
\frac{4\abs{\epsilonbold}^2}{E_+^2-E_-^2}\right)\tau_z\mathbbm{1}_4
\,{\clb \mp}\,
\frac{4i\omega_n\epsilon_0\epsilon_1}{E_+^2-E_-^2}\tau_0\gamma_1 - \epsilon_1\left(1 \,{\clb \pm}\,
\frac{4\epsilon_0^2}{E_+^2-E_-^2}\right)\tau_z\gamma_1
\right. \notag
\\
&&  \,{\clb \mp}\,
\frac{4i\omega_n\epsilon_0\epsilon_2}{E_+^2-E_-^2}\tau_0\gamma_2
{\clb \pm}\frac{4i\omega_n\Delta_0\epsilon_4}{E_+^2-E_-^2}\tau_x\gamma_2
- \epsilon_2\left(1 \,{\clb \pm}\, \frac{4(\epsilon_0^2 +
	\Delta_0^2)}{E_+^2-E_{-}^2}\right)\tau_z\gamma_2 - \epsilon_3\left(1 \,{\clb \pm}\, \frac{4(\epsilon_0^2 + \Delta_0^2)}{E_+^2-E_{-}^2}\right)\tau_0\gamma_3 
\notag \\
&& {\clb
	\pm}\frac{4\Delta_0\epsilon_1\epsilon_5}{E_+^2-E_-^2}\tau_x\gamma_3
{\clb \mp}
\frac{4i\omega_n\epsilon_0\epsilon_3}{E_+^2-E_-^2}\tau_z\gamma_3 {\clb
	\mp}
\frac{4i\omega_n\epsilon_0\epsilon_4}{E_+^2-E_-^2}\tau_0\gamma_4{\clb
	\mp} \frac{4i\omega_n\Delta_0\epsilon_2}{E_+^2-E_-^2}\tau_x\gamma_4
- \epsilon_4\left(1 \,{\clb \pm}\, \frac{4(\epsilon_0^2 +
	\Delta_0^2)}{E_+^2-E_{-}^2}\right)\tau_z\gamma_4\notag \\
&& - \epsilon_5\left(1 \,{\clb \pm}\, \frac{4(\epsilon_0^2 +
	\Delta_0^2)}{E_+^2-E_{-}^2}\right)\tau_0\gamma_5 {\clb
	\mp}\frac{4\Delta_0\epsilon_1\epsilon_3}{E_+^2-E_-^2}\tau_x\gamma_5 {\clb \mp}
\frac{4i\omega_n\epsilon_0\epsilon_5}{E_+^2-E_-^2}\tau_z\gamma_5 {\clb
	\pm}\frac{4\Delta_0\epsilon_1\epsilon_4}{E_+^2-E_-^2}\tau_yi\gamma_1\gamma_2 {\clb
	\pm}\frac{4i\omega_n\Delta_0\epsilon_5}{E_+^2-E_-^2}\tau_yi\gamma_1\gamma_3
\notag\\
&&\left. {\clb
	\mp}\frac{4\Delta_0\epsilon_1\epsilon_2}{E_+^2-E_-^2}\tau_yi\gamma_1\gamma_4 {\clb
	\mp}\frac{4i\omega_n\Delta_0\epsilon_3}{E_+^2-E_-^2}\tau_yi\gamma_1\gamma_5
- \Delta_0\left(1 \,{\clb \pm}\,
\frac{4(\abs{\epsilonbold}^2-\epsilon_1^2)}{E_+^2-E_{-}^2}\right)\tau_yi\gamma_2\gamma_4 {\clb
	\pm}\frac{4\Delta_0\epsilon_0\epsilon_1}{E_+^2-E_-^2}\tau_yi\gamma_3\gamma_5\right\}.
\end{eqnarray}
We have highlighted the symbols ${\clb \pm}$ and ${\clb \mp}$ in blue for clarity.

\bibliography{Quintet_Pairing_Ref}

\end{document}